\documentclass[aps,prd,twocolumn,superscriptaddress,showpacs]{revtex4-1}

\usepackage{amsmath}
\usepackage{amsfonts}
\usepackage{graphicx}

\usepackage{CJK}

\usepackage{color}

\begin{document}


\title{Quark Number Fluctuations at Finite Temperature and Finite Chemical Potential  via the Dyson-Schwinger Equation Approach}


\author{Xian-yin Xin }
\affiliation{Department of Physics and State Key Laboratory of Nuclear Physics and Technology, Peking University, Beijing 100871, China}

\author{Si-xue Qin }
\affiliation{Institute for Theoretical Physics, Johann Wolfgang Goethe University, D-60438 Frankfurt am Main, Germany}

\author{Yu-xin Liu }
\email[Corresponding author: ]{yxliu@pku.edu.cn} \affiliation{Department of Physics and State Key Laboratory of Nuclear Physics and Technology, Peking University, Beijing 100871, China} \affiliation{Collaborative Innovation Center of Quantum Matter, Beijing 100871, China} \affiliation{Center for High Energy Physics, Peking University, Beijing 100871, China}

\date{\today}
\begin{abstract}
We investigate the quark number fluctuations up to the fourth order in the matter composed of two light flavor quarks with isospin symmetry and at finite temperature and finite chemical potential using the Dyson-Schwinger equation approach of QCD. In order to solve the quark gap equation, we approximate the dressed quark-gluon vertex with the bare one and adopt both the Marris-Tandy (MT) model and the infrared constant (Qin-Chang) model for the dressed gluon propagator. Our results indicate that the second, third, and forth order fluctuations of net quark number all diverge at the critical end point (CEP). Around the CEP, the second order fluctuation possesses obvious pump while the third and fourth order ones exhibit distinct wiggles between positive and negative. For the MT model and the Qin-Chang model, we give the pseudo-critical temperature at zero quark chemical potential as $T_{c}=146$~MeV and $150$~MeV, and locate the CEP at $({\mu_{E}^{q}}, {T_{E}^{}}) = (120, 124)$~MeV and $(124,129)$~MeV, respectively. In addition, our results manifest that the fluctuations are insensitive to the details of the model, but the location of the CEP shifts to low chemical potential and high temperature as the confinement length scale increases.

\end{abstract}


\maketitle

\section{Introduction}

It is well known that the main objective of the study on QCD phase transitions is to make clear the phase structure of strong interaction matter at finite temperature and/or finite chemical potential~\cite{Gupta:2011Science,Fukushima:2011Review,Wambach:2009Review,Shuryak:2004PNPP}.  Since the critical end point (CEP) is a special state which separates the regions of first-order phase transition and crossover (or second-order phase transition), it becomes then the current focus of the theoretical and experimental investigations~\cite{Fukushima:2011Review,Gupta:2011Science,Melkumov:2012PAN,Odyniec:2012PAN}.  Many criteria, for instance, the disappearance of two minima of the thermodynamical potential, the chiral susceptibility, the disappearance of the S-shape relation between baryon chemical potential (related to the quark chemical potential simply with $\mu^{q} = \mu^{B}/3$) and baryon number density (related to the quark number density simply with $\rho^{q} = 3 \rho^{B}$), the fluctuations of conserved charges, finite-size scaling, thermal conductivity, and so on, have been proposed (see, for example, Refs.~\cite{Qin:2011PRL,Jiang:2013PRD,Liu:20101PRD,Stephanov:2009PRL,Stephanov:2011PRL,Zhou:2010PRC,Fraga:2011PRC,Kapusta:2012PRC}). Amongst them, the fluctuations can provide essential information about the effective degrees of freedom and their possible quasi-particle nature of the multi-particle system. It is also well known that the fluctuations are sensitive to the critical behaviors of the QCD matter, and thus can be taken as useful probes in exploring the QCD phase diagram, especially locating the CEP. What is more, the fluctuations can be extracted through experiments with event by event analysis~\cite{Stephanov:2009PRL,Zhou:2010PRC,Stephanov:1999PRD,Koch:20058PRLetc} which have been widely implemented in experimental researches (see, for instance, Refs.~\cite{Melkumov:2012PAN,Odyniec:2012PAN,Aggarwal:2010PRL,Tarnowsky:2011JPG,Luo:2012PAN,Mohanty:2013}), so they provide a direct connection between experiment and theory~\cite{Gupta:2011Science}.

On the theoretical side, great efforts have been made to investigate properties of the fluctuations. Lattice QCD simulations have given the temperature dependence of the fluctuations of conserved charges~\cite{Ejiri:2006PLB,Cheng:2009PRD,Bazavov:2012PRD1,Bazavov:2012PRD2,Gavai:2005PRD,Gavai:2008PRD,Aoki:2006PLB,Aoki:2009JHEP,Borsanyi:2010JHEP,Borsanyi:2012JHEP,Endrodi:2011JHEP,Allton:2003PRD,Allton:2005PRD,Bernard:20058PRD,Karsch:2011PRD}. The results show that the second order fluctuation experiences a rapid enhancement as the temperature increased and the higher order fluctuations exhibit a peak or a peak-valley structure which indicates that the system undergoes a continuous phase transition. However, the lattice QCD simulations cannot be extended to large chemical potential region in principle because of the ``sign problem''. In this case, continuum field theoretical analysis becomes a powerful tool which could work without such a handicap. Then plenty of works have been carried out in the Polyakov-loop extended Nambu--Jona-Lasinio (PNJL) model~\cite{Fukushima:2008PRD,Fu:2010PRD1,Fu:2010PRD2,Ray:20101PRD}, Polyakov-loop extended quark meson (PQM) model~\cite{Schaefer:20102PRD,Skokov:2011PRC,Mohan:2012PRD}, the hadron resonance gas (HRG) model~\cite{Huovinen:2010NPA,Cleymans:1999PRC,Andronic:2006NPA}, and other effective models~\cite{Stokic:2009PLB,Asakawa:2009PRL}. Moreover, the Dyson-Schwinger equation approach of QCD~\cite{Roberts:199400DSEint} has also been taken to calculate the quark number susceptibility~\cite{Zong:2009PRD,Zong:2009PLB,Zong:2011JHEP}, which is a measure of the quark number fluctuation. For the case at finite temperature and zero chemical potential, almost all these model calculations gave similar results as those obtained in lattice QCD simulations.

However, only quite limited works studied the case at both finite temperature and finite density (chemical potential)~\cite{Allton:2003PRD,Allton:2005PRD,Fu:2010PRD2,Skokov:2011PRC}. Based on these works, the researches of the fluctuations of the conserved charges at finite chemical potential are still quite far away from reaching the essential information of the CEP at present. In this sense, it is imperative to investigate this topic in continuum field theories, especially those with solid QCD foundations.

Dyson-Schwinger (DS) equations of QCD~\cite{Roberts:199400DSEint,Alkofer:2001DSE,Roberts:2003DSE,Roberts:20123Review} are believed to be one of the continuous field theory method that could describe both the dynamical chiral symmetry breaking (DCSB) and the confinement, simultaneously~\cite{McLerran:2007NPA}. It has produced many meaningful and instructive results on QCD phase transitions including the possibility of the existence and its location of the CEP~\cite{Qin:2011PRL,Zong:2009PRD,He:2007JPG,Fischer:20113PLB, Qin:2011PRD,Maris:2003EPJA,Qin:2013PRD,Fischer:2009PRL} and hadron structures~\cite{Roberts:20123Review,Chang:2009PRL,Qin:2011PRC,Qin:2012PRC,Eichmann:2010PRL,Eichmann:2011PRD,Chang:2013PRL,Wang:2013PRD}. Such a powerful approach has also been implemented to study the quark number susceptibility at finite temperature. The obtained results manifests that the DS equations can successfully describe the confinement and the DCSB at low temperature and the corresponding phase transitions with increasing temperature.

Inspired by the successes of the DS equation approach, we further study in this paper the quark number density fluctuations up to the fourth order in the matter, which is composed of up and down quarks with isospin symmetry, at not only finite temperature but also finite chemical potential, and locate the CEP of the QCD phase transitions. The obtained results are compared with those obtained in lattice QCD calculations and other phenomenological models. Especially, we translate the variation behaviors of the fluctuations into those of the experimental quantities which can be taken as the signatures of the CEP~\cite{Zhou:2010PRC,Luo:2012PAN,Mohanty:2013}. To show the universality of the results, we take both the Maris-Tandy (MT) model and the infrared constant (Qin-Chang, QC) model~\cite{Maris:1999PRC,Qin:2011PRC,Oliveira:2011JPG,Aguilar:2012PRD,Fischer:2012PRL,Dudal:2012PRD,Zwanziger:2013PRD} for the gluon propagator in our calculations.  Furthermore, we also discuss the effect of the model parameters on the phase diagram, especially, the location of the CEP.

This article is organized as follows. In Sec.~\ref{sec2}, we describe briefly the formalism of the thermodynamical fluctuations and the Dyson-Schwinger equations. In Sec.~\ref{sec3} we give our numerical results of the fluctuations up to the fourth order at finite temperature and finite chemical potential, and extract the QCD phase diagram and the location of the CEP. At the end of this section, the parameter dependence of the results is investigated. In Sec.~\ref{sec4}, we present our summary and some remarks.

\section{Dyson-Schwinger Equations approach}\label{sec2}

Consider a grand ensemble composed of particles whose chemical potential is $\mu_{X}^{}$, the partition function, or the generating functional in terms of the field operators $j_{0}^{} = \hat{n}(x) =\bar{q}(x){\gamma_{0}^{}}q(x) $ (the total particle number $ N_{X}:= \int {d^{4} x } \hat{n}(x) $) is usually written as
\begin{equation}
        \label{eq1}
        \begin{split}
                & \Xi [{\mu_{X}^{}} ; T] = \\
                & \int \mathcal{D}[\bar{q}, q, A_{\mu}, \bar{\omega}, \omega] \exp\left[-\beta \int {d^{4}x} (\mathcal{H}_{0} - \hat{n}(x) {\mu_{X}^{}} )\right] \, ,
        \end{split}
\end{equation}
where $\beta=1/T$ with $T$ being the temperature of the system. Then the so called grand thermodynamical potential reads,
\begin{equation}
        \Omega[\mu_X;T]=-\frac{1}{\beta}\ln\Xi[{\mu_{X}^{}};T].
\end{equation}
Thus the expectation value of the total particle number is obtained as
\begin{equation}
        \label{eq2} -\frac{\delta \Omega[{\mu_{X}^{}};T]}{\delta {\mu_{X}^{}} } = \int {d^{4} x } \langle \hat{n}(x) \rangle = \overline{N_{X}^{}} \, ,
\end{equation}
where $\langle\cdots\rangle$ and $\overline{\cdots}$ denote the ensemble averages. If doing the derivative of Eq.~(\ref{eq2}) over the chemical potential ${\mu_{X}^{}}$, we have
\begin{equation}
        \label{eq3} \frac{1}{\beta} \frac{
        \partial \overline{N_{X}^{}}}{
        \partial {\mu_{X}}} = \overline{N_{X}^{2}}-\overline{N_{X}^{}}^2 = \overline{({N_{X}^{}} - \overline{N_{X}^{}})^2} \equiv {\chi_{2}^{X}} V T^{3} \, .
\end{equation}
Namely, $\chi_{2}^{X}$ is related to the fluctuation of the conserved charge $\delta {N_{X}^{} }= {N_{X}^{}} - \overline{N_{X}^{}}$ as
\begin{equation}
        \label{eq4.1} \chi_{2}^{X} =\frac{1}{VT^{3}}\overline{\delta N_{X}^{2} } \, .
\end{equation}
Similarly, one could obtain the higher order fluctuations as follows
\begin{align}
        \chi_{3}^{X} & = \frac{1}{VT^3} \overline{\delta N_{X}^{3}} \, , \label{eq4.2} \\
        \chi_{4}^{X} & = \frac{1}{VT^3} (\overline{\delta N_{X}^{4} } - 3\overline{\delta N_{X}^{2} }^{2} ) \, , \label{eq4.3} \\
        \chi_{11}^{XY} & = \frac{1}{VT^3} \overline{\delta {N_{X}^{}} \delta {N_{Y}^{}} } \, , \label{eq4.4}
\end{align}
where $X$ and $Y$ stand for different conserved charges.

Generally, the thermodynamical potential can be expanded as the Taylor series of the chemical potentials of the conserved charges. The coefficients of the Taylor expansion correspond to the fluctuations, i.e.,
\begin{equation}
        \label{eq5} \chi^{XYZ\cdots}_{ijk\cdots} = -\frac{ T^{i+j+k+\cdots}}{V T^4}\frac{
        \partial^{i+j+k+\cdots}\Omega[ T; {\mu_{X}^{}} , {\mu_{Y}^{}} , {\mu_{Z}^{}} , \cdots ]}{
        \partial{\mu^i_{X}}
        \partial{\mu^j_{Y}}
        \partial{\mu^k_{Z}}\cdots} \, ,
\end{equation}
where ${\mu_{X}^{}},\; {\mu_{Y}^{}} , \; {\mu_{Z}^{}} , \cdots$ stands for the chemical potential of the conserved charge $X,\; Y,\; Z, \cdots$, respectively. In what follows, since we only study the quark number fluctuations, the superscript of $\chi$ is suppressed unless stated. In principle, one can derive the fluctuations according to Eq.~(\ref{eq5}) if the thermodynamical potential is available. However, sometimes one could not have the exact expression of the thermodynamical potential due to the complexity of the non-perurbative QCD. In this case, as we will see the particle number can still be easily calculated, thus one can start with Eq.~(\ref{eq2}), equivalently.

The second order fluctuation ${\chi_{2}^{}}$ can be taken as an effective criterion of the phase transition similar with the chiral susceptibility $\chi_c$ which is defined as the derivative of the chiral condensate $\langle \bar{q} q \rangle$ to the current quark mass $m$, i.e., $\chi_{c} = - \frac{
\partial \langle \bar{q} q \rangle}{
\partial m} $ . According to Eq.~(\ref{eq4.1}), it is expected that $\chi_{2}^{}$ goes to divergence if the system undergoes a first-order phase transition. Moreover, the correlation length $\xi$ also diverges since it can be related to $\chi_2$, i.e., $\chi_2 \sim \xi^{2-\eta}$ where $\eta$ is a small and positive number~\cite{Stephanov:2004APPB}.

In this work, we consider a quark system composed of $u$ and $d$ quarks with exact isospin symmetry, and set $\mu_{u}^{} = \mu_{d}^{} = \mu$ for simplicity. We suppose that the system is uniform, i.e., $\overline{N} = V n$ with $n$ being the particle number density. Then, the fluctuations could be derived from the certain order derivatives of $n(\mu,T)$. In terms of the quark propagator, we can derive the expression of $n(\mu,T)$ as
\begin{align}
        n(\mu,T) & = 2N_{c} N_{f} Z_{2} \int_{-\infty}^{\infty} \frac{{d^{3}} \vec{p}}{(2\pi)^3}{f_{1}^{}}(\vec{p};\mu,T) \, , \label{eq8} \\
        {f_{1}^{}}(\vec{p};\mu,T) & = \frac{T}{2} \sum_{m=-\infty}^{\infty} {\rm tr_{D}^{}}(-{\gamma_{4}^{}} S(\tilde{\omega}_{m}^{}, \vec{p}\,)) \, , \label{eq9}
\end{align}
where $Z_2$ is the quark wave-function normalization constant, $N_c=3$ is the color number, $N_f=2$ is the flavor number, $S(\tilde{\omega}_{m}^{}, \vec{p}\,)$ is the quark propagator, and the summation is taken over the Matsubara frequencies ($\tilde{\omega}_{m}^{} =\omega_m+i\mu= (2m+1)\pi T + i \mu$, $ m\in \mathbb{Z}$). According to its Lorentz structure, the quark propagator can be expressed as
\begin{equation}
        \label{eq10}
        \begin{split}
                S(\tilde{\omega}_{m}, \vec{p}\,)^{-1} & = i \vec{\gamma}\cdot \vec{p} \, A(\tilde{\omega}_{m}^{2},\vec{p}\,^2) \\
                & \quad + i {\gamma_{4}^{}} \tilde{\omega}_{m}^{} C(\tilde{\omega}_{m}^{2}, \vec{p}\,^2) + B(\tilde{\omega}_{m}^{2}, \vec{p}\,^2) \, ,
        \end{split}
\end{equation}
where $A$, $B$ and $C$ are scalar functions. Thus we can rewrite ${f_{1}^{}}(\vec{p};\mu,T)$ in terms of the scalar functions $A$, $B$, and $C$ as
\begin{equation}
        \label{eq11}
        \begin{split}
                {f_{1}^{}}(\vec{p};\mu,T)= T \sum_{m=-\infty}^{\infty} \Big[\frac{1/C}{\mu+E-i {\omega}_{m}^{} } + \frac{1/C}{\mu-E-i {\omega}_{m}^{} } \Big] \, ,
        \end{split}
\end{equation}
where $E=\sqrt{\vec{p}\,^{2} (A/C)^{2} + (B/C)^{2}}$. For ideal quark gas, i.e., $A=C=1$ and $B$ equals to the quark mass, Eq.~(\ref{eq11}) is exactly the sum of the quark and anti-quark Fermi distributions~\cite{He:2007JPG}. For the quark system with interaction, the scalar functions $A$, $B,$ and $C$ are modified from that of the idea quark gas. These scalar functions can be determined by solving the gap equation, i.e., the Dyson-Schwinger equation for the quark propagator,
\begin{gather}
        S(\tilde{\omega}_{n}^{},\vec{p}\,)^{-1}=i \vec{\gamma}\cdot \vec{p} + i {\gamma_{4}^{}} \tilde{\omega}_{n}^{} + m_{0} + \Sigma(\tilde{\omega}_{n}^{} , \vec{p}\,) \, , \label{eq12} \\
        \begin{split}
                \label{eq13} \Sigma(\tilde{\omega}_{n}^{}, \vec{p}\,) & = T\sum_{m=-\infty}^{\infty}\int \frac{{d^{3}} \vec{q}}{(2\pi)^{3}} {g^{2}} D_{\mu\nu}(\vec{p} - \vec{q}, \Omega_{nm}; T, \mu) \\
                & \quad \times \frac{\lambda^{a}}{2} {\gamma_{\mu}^{}} S(\tilde{\omega}_{m}^{} , \vec{q}) \frac{\lambda^{a}}{2} \Gamma_{\nu}(\tilde{\omega}_{m}^{}, \vec{q}, \tilde{\omega}_{n}^{}, \vec{p}\,) \, ,
        \end{split}
\end{gather}
where $\Omega_{nm}=\omega_{n}^{} - \omega_{m}^{}$, $D_{\mu\nu}$ is the dressed-gluon propagator, and $\Gamma_{\nu}$ is the dressed quark-gluon vertex. The gap equation is closed if the dressed gluon propagator and the dressed quark-gluon vertex are specified. In this work, we take the rainbow approximation, i.e.,
\begin{equation}
        \label{eq14} \Gamma_{\nu}(\tilde{\omega}_{m}^{}, \vec{q}, \tilde{\omega}_{n}^{}, \vec{p}\,) = {\gamma_{\nu}^{}} \, .
\end{equation}
For the dressed gluon propagator, one can solve its DS equation which couples with the quark's gap equation and involves the 3-gluon vertex, the 4-gluon vertex, and the ghost. To solve these coupled and unclosed equations, one needs a subtle truncation scheme which makes the calculation very complicated. After some simplification, e.g., neglecting the gluon vertex and the ghost, solving the coupled equations becomes feasible~\cite{Fischer:20113PLB,Luecker:2013PoS}. However, in this work, we adopt the effective model which keeps the main feature of the gluon propagator and makes the calculation much simpler.

Formally, the dressed gluon propagator can be decomposed as the transverse and longitudinal parts, i.e.,
\begin{equation}
        \label{eq15} g^{2} D_{\mu\nu}(\vec{k}, \Omega_{nm}) = P^{T}_{\mu\nu} D_{T}(\vec{k}^{2}, \Omega^{2}_{nm}) + P^{L}_{\mu\nu} D_{L}(\vec{k}^{2}, \Omega^{2}_{nm}) \, ,
\end{equation}
where $P_{\mu\nu}^{T,L}$ are the transverse and longitudinal projection operators, respectively, as
\begin{align}
        P_{\mu\nu}^{T} & =
        \begin{cases}
                0, & \mu \quad \textrm{and/or} \quad \nu=4, \\
                \delta_{ij}-\frac{\vec{k}_i\vec{k}_j}{\vec{k}^2}, & \mu,\nu=1,2,3,
        \end{cases}
        \label{eq16} \\
        P_{\mu\nu}^{L} & = \delta_{\mu\nu} - \frac{{k}_{\mu}{k}_{\nu}}{k^2} - P_{\mu\nu}^{T} \, , \label{eq17}
\end{align}
and the $D_{T,L}$ are the gluon dressing functions,
\begin{align}
        D_{T} & = \mathcal{D}(\vec{k}^{2} + \Omega_{nm}^{2}) \, , \label{eq18p1} \\
        D_{L} & = \mathcal{D}(\vec{k}^{2} + \Omega_{nm}^{2} +m_{g}^{2}) \, , \label{eq18p2}
\end{align}
where $m_{g}^{}$ is the gluon Debye mass which depends on temperature, i.e., $m_{g}^{2}(T)=(16/5) T^{2}$~\cite{Aguilar:2012PRD,Kondo:1997IJMPA,Qin:2011PRD}. The scalar function is modeled as two parts,
\begin{equation}
        \label{eq:mdl} \mathcal{D}(s)=\mathcal{D}_{\rm IR}(s) + 4 \pi \mathcal{F}(s){\alpha_{\rm pQCD}^{}} (s) \, .
\end{equation}
The second part expresses the one-loop perturbative result which dominates the ultraviolet behavior of the interaction,
\begin{equation}
        \label{eq19} \mathcal{F}(s)\alpha_{\rm pQCD}^{}(s) = \frac{2\pi {\gamma_{m}^{}}(1-e^{-s/4 m_{t}^{2}})}{\ln[\tau+(1+s/\Lambda^{2}_{QCD})^{2}]} \, ,
\end{equation}
where $\tau=e^{2} - 1$, $m_{t}^{} = 0.5$~GeV, the anomalous dimension $\gamma_{m}^{} = 12/(33-2N_{f})$ with $N_{f}=4$, and the QCD scale $\Lambda_{\rm QCD}^{N_{f}=4} = 0.234$~GeV. The infrared part can have many forms. In this work, we take two typical ones, i.e., the Marris-Tandy (MT) model~\cite{Maris:1999PRC} and the Qin-Chang (QC) model~\cite{Qin:2011PRC}, which read
\begin{align}
        \mathcal{D}_{\rm IR}^{\rm MT}(s)&=\frac{4\pi^2D}{\omega^6}se^{-s/\omega^2} \, , \label{eq20}\\
        \mathcal{D}_{\rm IR}^{\rm QC}(s)&=\frac{8\pi^2D}{\omega^4}e^{-s/\omega^2} \, , \label{eq21}
\end{align}
where $\omega$ and $D$ are the interaction width and strength, respectively. Note that in the infrared region the MT model vanishes, while the QC one goes to a constant. Recent lattice QCD simulations and gauge field theory analysis prefer the QC model~\cite{Oliveira:2011JPG,Aguilar:2012PRD,Fischer:2012PRL,Dudal:2012PRD,Zwanziger:2013PRD}. However, we consider both the two forms and look through the model dependence in this work.

To make the model more practical, we consider the temperature dependance of the infrared interaction strength. Note that the temperature serves a energy scale to suppress the infrared interaction because of the running behavior of the QCD coupling constant. So it is natural that the infrared strength of the interaction decreases logarithmically when the temperature is high enough. Following Ref.~\cite{Qin:2011PRD}, we take a simple ansatz as
\begin{equation}
        \label{eq22} D(T)=
        \begin{cases}
                D\, ,  & T< T_{p} \, , \\
                \frac{a}{b+\ln[T/\Lambda_{QCD}]} \, , & T>T_{p} \, ,
        \end{cases}
\end{equation}
where $T_{p}$ is a scale below which nonperturbative effects associated with confinement and DCSB are not materially influenced by the thermal screening~\cite{Qin:2011PRD}. The parameters $a$ and $b$ are determined by two constraints: $D(T_{p}^{})=D$
at $T_{p}$; the quark thermal mass satisfies $m_{T}^{} = 0.8T$ when $T\gtrsim 2T_{c}$, which is consistence with the results given in lattice QCD simulation \cite{Karsch:2009PRD}. Herein, we set $T_{p}=1.3\, T_{c}$ at which the strong correlation among quarks decreases drastically~\cite{Qin:2011PRD}.

The two main parameters $D$ and $\omega$ in our models are chosen as those reproducing the masses and form factors of $\pi$, $\rho$, $K$, $\phi$ and $\omega$ mesons in vacuum. Calculations show that these observables are insensitive to the variation of $\omega \in [0.3, 0.5] $~GeV in the MT model~\cite{Maris:2003EPJA} and $\omega \in [0.4, 0.6] $~GeV in the QC model~\cite{Qin:2011PRC} as long as $D\omega={\rm const}$. Following Ref.~\cite{Qin:2011PRC}, we choose $(D\omega)^{1/3}=0.72$ with $\omega = 0.4$ GeV for MT model and $(D\omega)^{1/3}=0.80$ with $\omega = 0.5$ GeV for QC model. We will also analyze how the parameter $\omega$ affects the results. The renormalization point is fixed at $\zeta =19$~GeV as Refs.~\cite{Maris:1997PRC,Maris:1999PRC}, and the corresponding current quark mass is taken as $m_{u}^{} =m_{d}^{} =5.5$~MeV.

\section{Numerical Results and Discussion}\label{sec3}

\begin{figure}
        \includegraphics[width=0.5\textwidth]{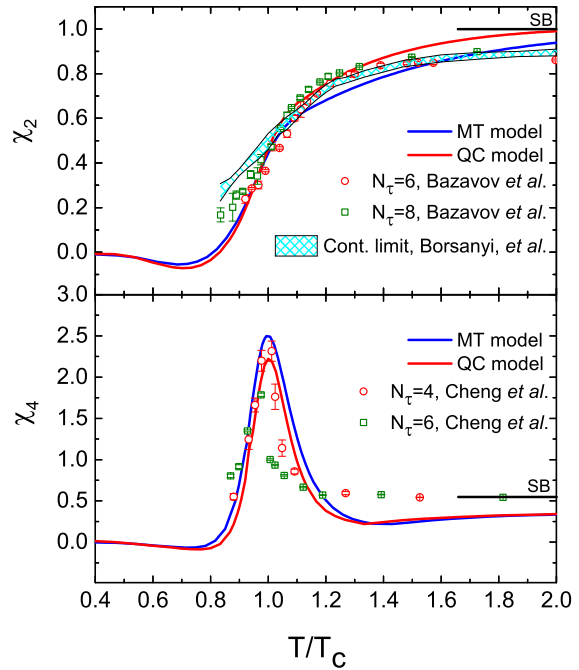}
        \caption{(color online) The second and fourth order fluctuations in the MT model with $\omega^{\rm MT}=0.4$~GeV and the QC model with $\omega^{\rm QC}=0.5$~GeV. The lattice QCD results (taken from Refs.~\cite{Cheng:2009PRD,Bazavov:2012PRD1,Borsanyi:2012JHEP} and denoted by data points with error bars) are also displayed for comparison.} \label{figT}
\end{figure}

Fig.~\ref{figT} illustrates the calculated second and fourth order quark number fluctuations at zero quark chemical potential and finite temperature. It is noticed that the second order fluctuation experiences a rapid growth as the temperature increases. Its derivative with respect to the temperature exhibits exactly the same appearance of the fourth order fluctuation shown in the lower panel of Fig.~\ref{figT}. The temperature for the derivative or the fourth order fluctuation to take its maximum is $146~(150)$~MeV in the MT (QC) model. We obtain then the pseudo-critical temperature $T_{c}=146~(150)$~MeV in the MT (QC) model. Lattice QCD gave different pseudo-critical temperatures with different calculation schemes, typically, $T_{c} \in [140,160]$~MeV for light quark system (see, for example, Refs.~\cite{Bazavov:2012PRD1,Bazavov:2012PRD2,Aoki:2006PLB,Aoki:2009JHEP,Borsanyi:2010JHEP,Borsanyi:2012JHEP}). It is apparent that our results of the pseudo-critical temperature agree nicely with those obtained in lattice QCD. Moreover, the calculated temperature dependent behaviors of the fluctuations also agree with those obtained in lattice QCD \cite{Cheng:2009PRD,Bazavov:2012PRD1,Borsanyi:2012JHEP} (marked by data points in Fig.~\ref{figT}) in a quite large domain, especially, around the pseudo-critical temperature.

One can also read from the figure that, as the temperature gets very high, the second order fluctuation approaches the Stefan--Boltzmann (SB) limit, i.e., $\chi_{2}^{\rm SB}=1$, and the fourth order one decreases to a small value. Moreover, the results with the two models of the gluon propagator give almost the same result. Such features can be easily understood as follows. According to Eq.~\eqref{eq22}, the infrared interaction strength is screened at high temperature so that only the ultraviolet part dominates the interaction. As a consequence, the two models actually degenerate into the same form derived from perturbation theory
(see Eq.~\eqref{eq:mdl}).

Another characteristic of the fluctuations shown in Fig.~\ref{figT} is that both the second and fourth order fluctuations at lower temperature are negative with small magnitudes. At zero chemical potential, the second order fluctuation is actually the quark number susceptibility which measures the response of the quark number density to quark chemical potential. A negative value means that the quark number density is negative under the perturbation of the external source, or in other word, the external source is impossible to excite quarks. Thus, such a feature of the fluctuation at low temperature is related to the quark confinement. On the other hand, the negative fluctuations can be understood from the viewpoint of the quark spectral function. The axiom of reflection positivity states that the propagator for an asymptotic quark must have a positive definite K\"{a}llen-Lehmann spectral representation. The positivity violation of the quark spectral function means that quarks have to be somehow confined, and thus serves as a sufficient condition for quark confinement~\cite{Qin:2013PRD,Hawes:1994PRD}. According to Eqs. \eqref{eq8} and \eqref{eq9}, the quark number density can be calculated by the quark propagator. Then, using the quark spectral representation, the fluctuations can be expressed from the quark spectral function. Actually, the negative fluctuations reflect the non-positive definite quark spectral function. Namely, the system is confined at low temperature.

\begin{figure}
        \includegraphics[width=0.5\textwidth]{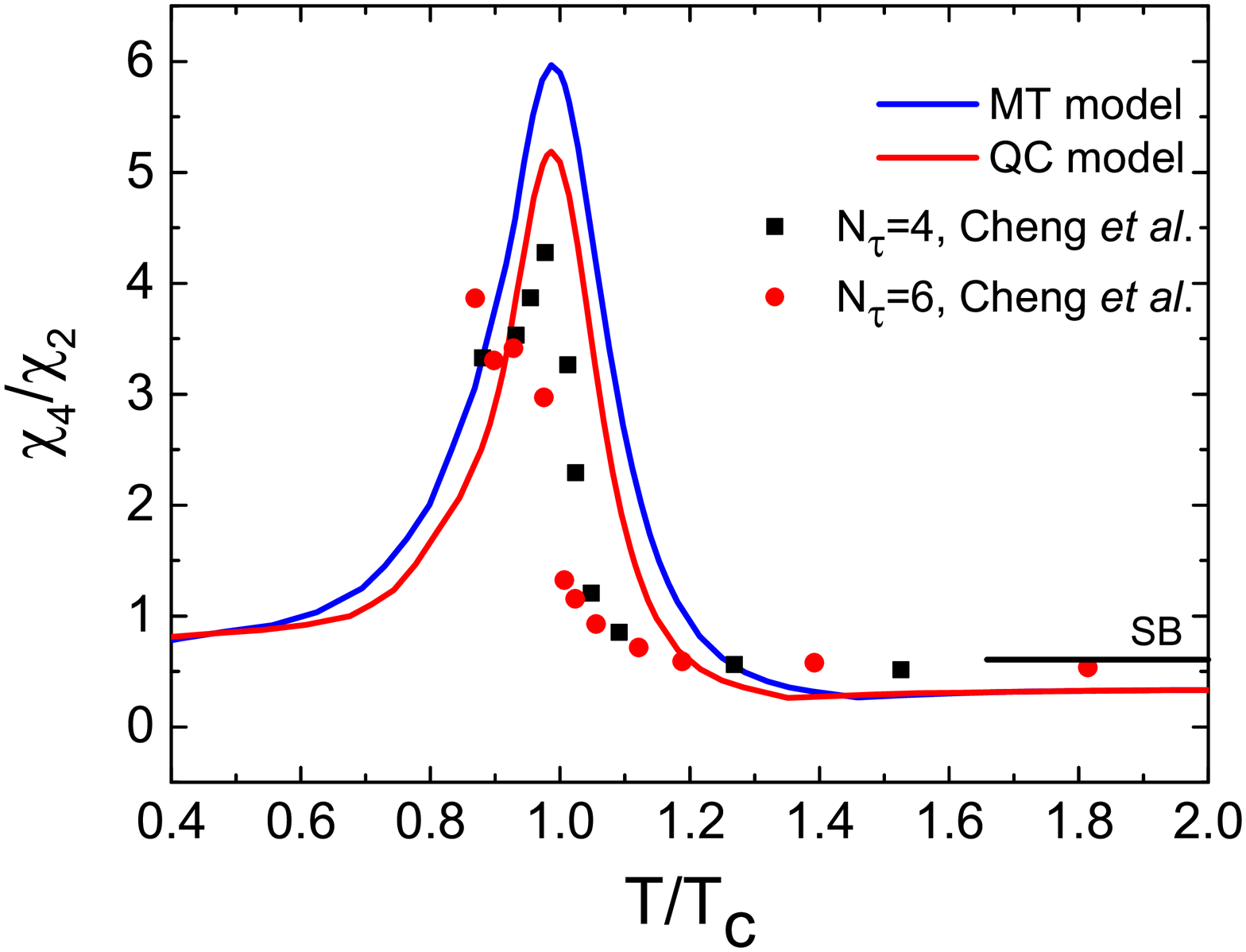}
        \caption{(color online) The temperature dependence of the ratio of the second to fourth order fluctuations in the MT model with $\omega^{\rm MT}=0.4$~GeV and the QC model with $\omega^{\rm QC}=0.5$~GeV. For comparison, the lattice QCD results (taken from Ref.~\cite{Cheng:2009PRD}) are denoted by data points.} \label{figTchi42}
\end{figure}

Next we discuss the fluctuation ratios which are very important because they are related to the experimental statistical variance, skewness, and kurtosis (denoted by $\sigma$, $S$, and $\kappa$, respectively). Explicitly, we have
\begin{equation}
        \label{eq:exps} S \sigma = \frac{\chi_3}{\chi_2}, \qquad \kappa \sigma^{2} = \frac{\chi_4}{\chi_2} \, .
\end{equation}
So the ratios can serve as a bridge connecting theoretical results and experimental data~\cite{Gupta:2011Science,Zhou:2010PRC,Luo:2012PAN,Mohanty:2013,Ejiri:2006PLB,Stokic:2009PLB}. Another advantage of the ratios is that they cancel the volume effects. The obtained temperature dependence of ${\chi_{4}^{}}/{\chi_{2}^{}}$ at zero chemical potential is shown in Fig.~\ref{figTchi42}, where lattice QCD results are also presented as a comparison. It can be noticed that with increasing temperature the ratio ${\chi_{4}^{}}/{\chi_{2}^{}}$ increases for $T<T_{c}$ while decreases for $T>T_{c}$. At very high temperature, the ratio goes to the SB limit. Meanwhile, the ratios given by the two interaction models exhibit the same behavior which is consistent with that obtained in lattice QCD.

\begin{figure}
        \includegraphics[width=0.5\textwidth]{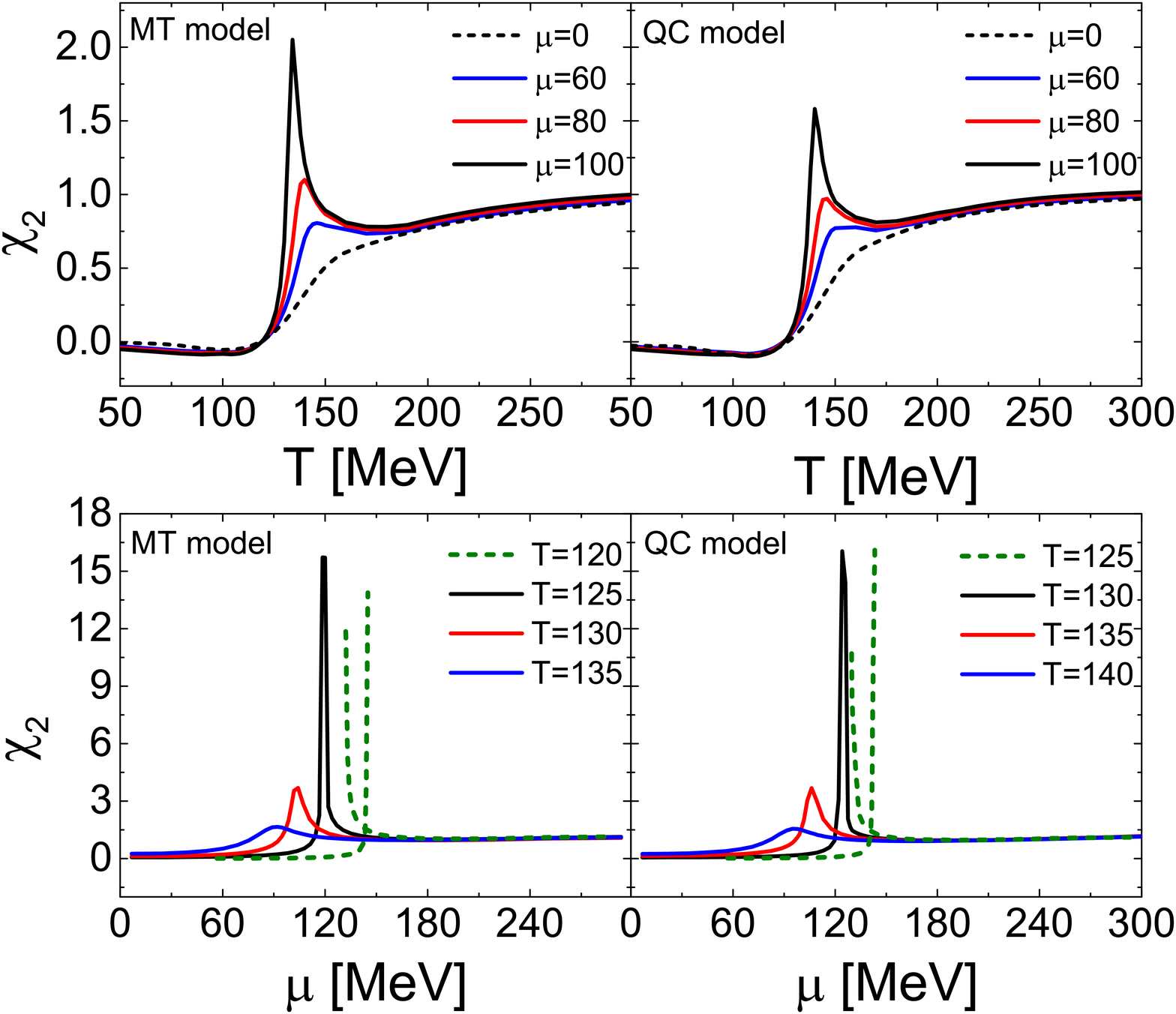}
        \caption{(color online) The variation behaviors of the second order fluctuation with respect to temperature at several values of the quark chemical potentials (upper panels) and those with respect to quark chemical potential at several values of the temperature (lower panels). For the two cases, the left panel shows the result in the MT model with $\omega^{\rm MT}=0.4$~GeV, and the right panel in the QC model with $\omega^{\rm QC}=0.5$~GeV. } \label{figTmu}
\end{figure}

The above results illustrate that the DS equation approach is as powerful to describe the quark number fluctuations at finite temperature and zero chemical potential as lattice QCD and effective models (see, for example, Refs.~\cite{Fu:2010PRD1,Schaefer:20102PRD,Mohan:2012PRD}). We then extend our calculations to the cases at finite temperature and finite chemical potential. Fig.~\ref{figTmu} displays the variation behaviors of the second order fluctuation with respect to temperature and chemical potential in different conditions. The upper panels of Fig.~\ref{figTmu} show that for $\mu>0$ the fluctuation $\chi_{2}^{}$ exhibits a bulge which grows and becomes a peak with increasing chemical potential. Meanwhile, the peak goes sharper and shifts to lower temperature. At high chemical potential, the peak evolves into a singularity. This behavior is consistent with that given in the effective model~\cite{Fu:2010PRD2}. From the lower panels of Fig.~\ref{figTmu}, one can recognize that $\chi_{2}^{}$ goes to nearly-divergent at $\mu=120~(124)$~MeV when $T=125~(130)$~MeV in the MT (QC) model. At higher temperature, the very sharp peak smears to a wide hump locating at lower chemical potential. Whereas, at lower temperature, e.g., $T=120~(125)$~MeV for the MT (QC) model, $\chi_{2}^{}$ gains two separate divergent points at different chemical potentials (denoted by dashed lines in the lower panels of Fig.~\ref{figTmu}). In such a case, $\chi_{2}^{}$ breaks into the left and right branches which extend to the low and high chemical potential regions, respectively. In the low chemical potential region, the system is characterized by DCSB and thus stays in the Nambu phase. In the high chemical potential region, the chiral symmetry is restored, and the system thus stays in the Wigner phase. It is noticed that the two branches overlap with each other in the region $\mu\in[132,143]$~MeV for the MT model. In the domain, the Nambu and Wigner phases coexist, which means that the phase transition is a first order one. Recalling that the system undergoes crossover at low chemical potential and high temperature, we find that there exists a CEP which separates the crossover and the first order phase transition in the $T$--$\mu$ plane. Then, we obtain the CEP locating at $(\mu_{E}^{q} ,T_{E}^{}) = (120, 124)$ MeV and $(124, 129)$~MeV in the MT and QC model, respectively.

\begin{figure}
        \includegraphics[width=0.5\textwidth]{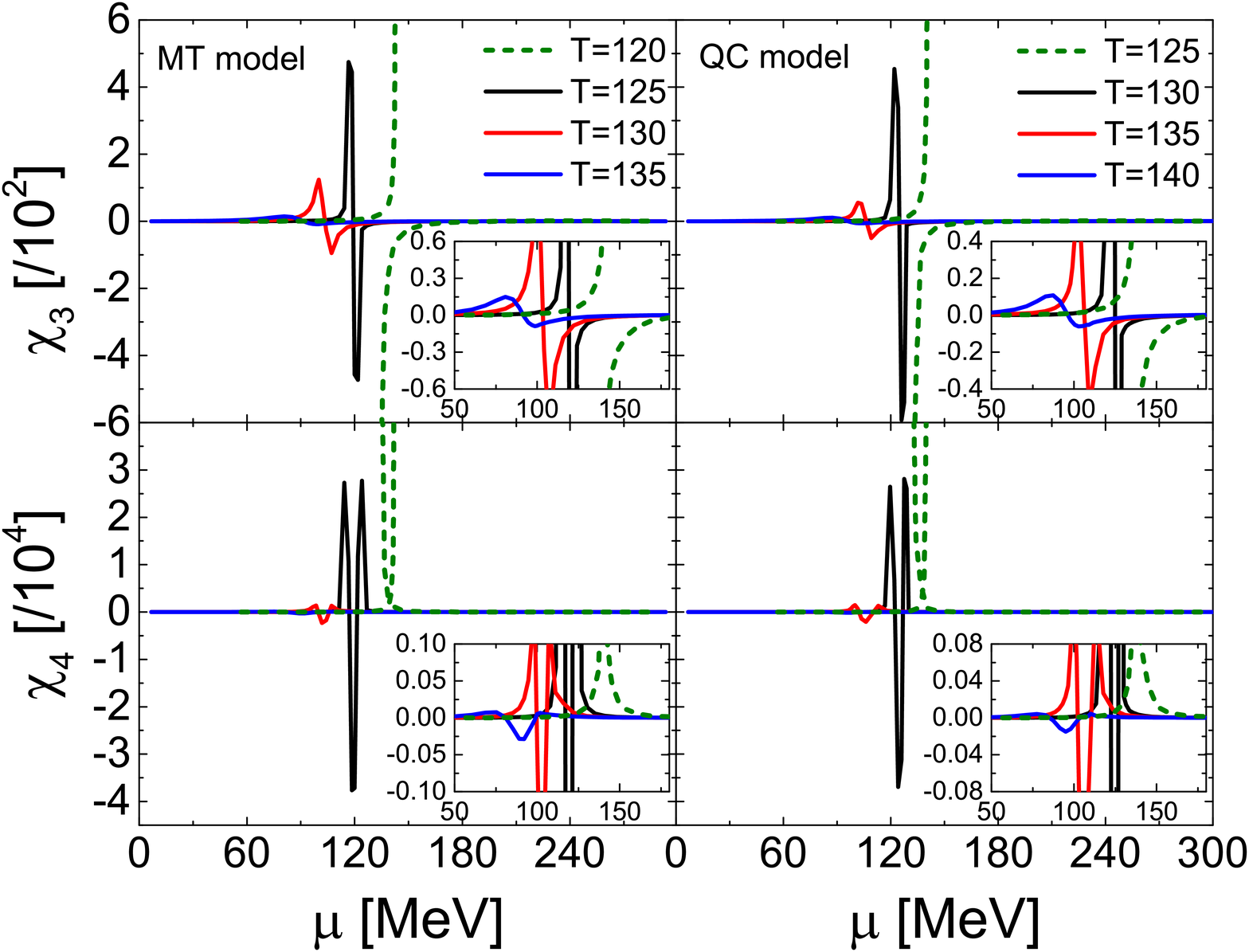}
        \caption{(color online) The variation behaviors of the third (upper panels) and fourth (lower panels) order fluctuations against quark chemical potential at several temperatures. The left panels show the results in the MT model with $\omega^{\rm MT}=0.4$~GeV, and the right panels in the QC model with $\omega^{\rm QC}=0.5$~GeV.} \label{figmu}
\end{figure}

It is believed that the higher order fluctuations are more sensitive to the critical behaviors of the QCD thermodynamics~\cite{Zhou:2010PRC,Stephanov:2011PRL,Fraga:2011PRC,Kapusta:2012PRC}. Fig.~\ref{figmu} shows the variation behaviors of the third and fourth order quark number fluctuations with respect to quark chemical potential at several temperatures. From the upper panels, we find that the third order fluctuation changes its sign at the critical point, which is consistent with the results given in Refs.~\cite{Stephanov:2011PRL,Fu:2010PRD2}. Since $\chi_4$ can be expressed as the derivative of $\chi_3$ to chemical potential, the global minimum of $\chi_4$ corresponds to the zero (or sign-changing) point of $\chi_3$  (compare the lower panels with the upper ones in Fig.~\ref{figmu}). As the temperature goes closer to that of the CEP, the wiggles of the fluctuations becomes sharper. At the CEP, both $\chi_{3}^{}$ and $\chi_{4}^{}$ diverge. Thus, tracking their singularity locations, we can locate the CEP. For $T<T_{E}^{}$, e.g., $T=120~(125)$ MeV in the MT (QC) model, both $\chi_3$ and $\chi_{4}^{}$ possess two separate singular points at the same chemical potentials as $\chi_{2}^{}$ (see the dash lines in Fig.~\ref{figmu}). According to Eq.~\eqref{eq5}, this can be easily understood because the higher fluctuation has to diverge if the lower one diverges at some temperature and chemical potential.
\begin{figure}
        \includegraphics[width=0.5\textwidth]{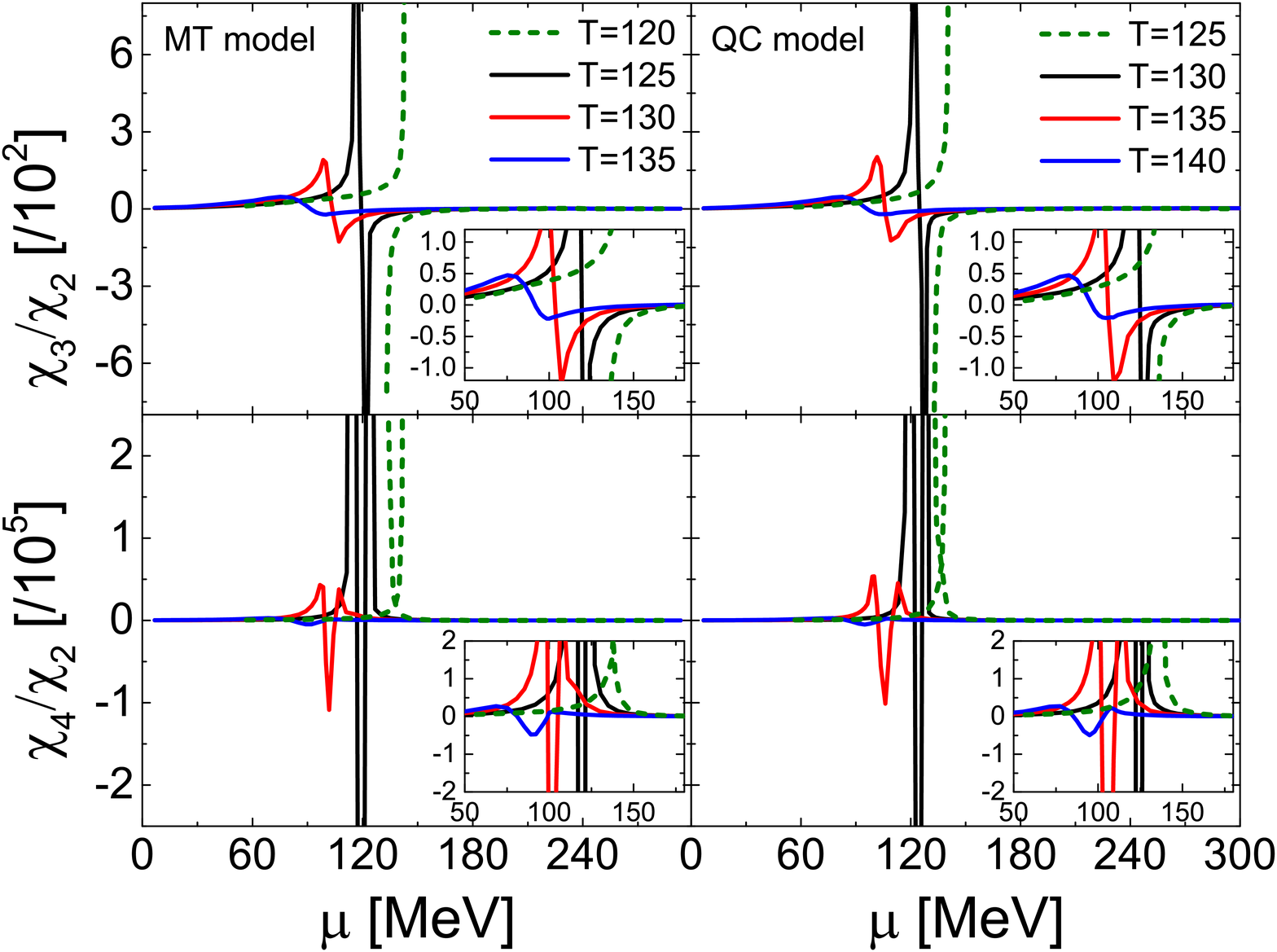}
        \caption{(color online) The quark chemical potential dependence of the ratio of the third to second order fluctuations at several temperatures (upper panels) and that of the fourth to second order fluctuations (lower panels). The left panels show the results in the MT model with $\omega^{\rm MT}=0.4$~GeV, and the right panels in the QC model with $\omega^{\rm QC}=0.5$~GeV. } \label{figmu3242}
\end{figure}

Fig.~\ref{figmu3242} displays the variation behaviors of the ratios of the high to low order fluctuations, e.g., $\chi_3/\chi_2$ and $\chi_4/\chi_2$. It is found that the ratios behave as the same as the third and fourth order fluctuations do (shown in Fig.~\ref{figmu}). Nevertheless, $\chi_3/\chi_2$ and $\chi_4/\chi_2$ have much larger amplitudes than $\chi_3$ and $\chi_4$, which means that the experimental quantities (see Eq. \eqref{eq:exps}) are very sensitive to the critical behaviors. As we know that the divergence of the fluctuations in thermal equilibrium at the critical point arises from the divergence of the correlation length~\cite{Stephanov:2004APPB}, while our calculation has not yet involved any non-equilibrium effects. In fact, the critical slowing down effect can reduce the growth of the correlation length, and in turn the fluctuations, in experiments (see {\it e.g.} Ref.~\cite{Rajagopal:2000PRD}). Therefore, the divergences of these quantities at the CEP state may be smeared into sharp peaks in experiments and can then be taken as practical  signatures of the CEP.

\begin{figure}
        \includegraphics[width=0.5\textwidth]{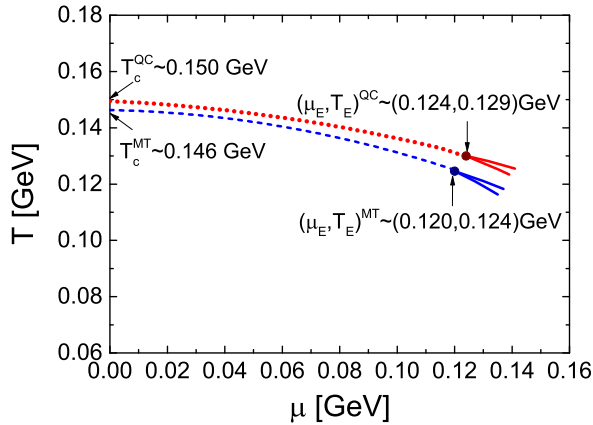}
        \caption{(color online) The QCD phase diagram obtained with the fluctuation criterion. The lower curves stand for the result in the MT model with $\omega^{\rm MT}=0.4$~GeV, and the upper curves in the QC model with $\omega^{\rm QC}=0.5$~GeV. } \label{figPhD}
\end{figure}

Analyzing the characteristics of the fluctuations in the $T$--$\mu$ plane, we can chart the QCD phase diagram. The obtained results in the two interaction models are illustrated in Fig.~\ref{figPhD}. It is shown that the system undergoes crossover in high temperature and low chemical potential region, while first order phase transition in the low temperature and high chemical potential region. This means that there exists a CEP in the $T$--$\mu$ plane. The pseudo-critical temperature at zero quark chemical potential is $T_{c}(\mu^{q}\! = \! 0)=146~(150)$~MeV in the MT (QC) model. The CEP locates at $({\mu_{E}^{q}} , {T_{E}^{}})=(120,124)$ MeV and $(124,129)$~MeV in the MT and QC model, respectively. It is noticed that the two models both give ${\mu_{E}^{q}}/{T_{E}^{}} \approx 1$, which agrees with that obtained by analyzing the chiral susceptibility in the DS equation approach~\cite{Qin:2011PRL}. Moreover, this result is also consistent with those given in other approaches, e.g., the elliptic flow data analysis~\cite{Lacey:2007na}  and the non-local NJL model~\cite{Weise:200910PRD,Contrera:2010}. Besides, although lattice QCD calculations have not yet given certain conclusion on the location or even the existence of the CEP~\cite{Endrodi:2011JHEP}, the ones with reweighting technique~\cite{Fodor:2002JHEP} or Taylor expansion method~\cite{Schmidt:2008JPG} or canonical ensemble approach~\cite{Liu:20101PRD} gave the similar results, and others  with reweighting technique~\cite{Fodor:2004JHEP}  or Taylor expansion method~\cite{Gavai:2005PRD,Gavai:2008PRD,Gupta:2014} gave smaller values, i.e., ${\mu_{E}^{q}}/{T_{E}^{}} \sim (0.4,0.7)$.

In our calculations, we adopt two different models, i.e., the MT model and the QC model, for the dressed gluon propagator. The above description indicates that the results obtained in the two models are very close to each other except for some quantitative differences. This can be understood as follows. Both the MT model and the QC model preserve the one-loop renormalization-group behavior of QCD~\cite{Qin:2011PRC} (see Eq.~\eqref{eq:mdl}). From Eqs.~(\ref{eq20}) and (\ref{eq21}) we can see that the main difference between the two models lies only in the deep infrared domain: The MT model vanishes when the gluon momentum goes to zero while the QC one to a finite constant. However, because of the energy scale resulted from the finite temperature, the dressed gluon propagator in the deep infrared domain has a very week affect on the effective interaction. Thus, the behaviors of the system become insensitive to the model details in the deep infrared domain.

With the constraint $D\omega = {\rm const}$, the two models still have a free parameter $\omega$. It is then interesting to investigate the effect of the parameter on the phase diagram. We have then performed calculations with various values of $\omega$. Since the phase diagram can be featured with the pseudo-critical temperature $T_{c}^{}$ and the location of the CEP (see Fig.~\ref{figPhD}), we list these quantities in Table~\ref{Table-sigma}. It is found that, with increasing $\omega$, the pseudo-critical temperature $T_{c}^{}$ gets lower and the ratio ${\mu_{E}^{q}}/{T_{E}^{}}$ becomes larger with the CEP rotating to the $\mu$-axis. This is consistent with that given by analyzing the chiral susceptibility in the DS equation approach~\cite{Qin:2011PRL}.
\begin{table}
        \caption{The pseudo-critical temperature $T_{c}^{}$ and the location of the CEP $({\mu_{E}^{q}}, {T_{E}^{}})$ calculated with several values of the parameter $\omega$ (dimensional quantities are given in GeV).}
        \begin{tabular}
                {c|ccccc} \hline \hline Model & ~$(D\omega)^{1/3} $~ & $\omega$ & $T_{c} $ & $({\mu_{E}^{q}}, {T_{E}^{}})$ & $\mu_{E}^{q}/T_E$ \\
                \hline {} & $0.72$ & ~$0.40$~ & ~$0.146$~ & ~$(0.120, 0.124)$~ & 0.97 \\
                MT~ & $0.72$ & $0.45$ & $0.132$ & $(0.220, 0.098) $ & 2.24 \\
                ~ & $0.72$ & $0.50$ & $0.124$ & $(0.281, 0.070) $ & 4.01 \\
                \hline {} & $0.80$ & $0.40$ & $0.173$ & $(0.075, 0.165) $ & 0.45 \\
                QC~ & $0.80$ & $0.50$ & $0.150$ & $(0.124, 0.129) $ & 0.96 \\
                ~ & $0.80$ & $0.60$ & $0.131$ & $(0.201, 0.099) $ & 2.03 \\
                \hline \hline
        \end{tabular}
        \label{Table-sigma}
\end{table}
It is known that the parameter $\omega$ characterizes the interaction width in the momentum space. Then, one can introduce a confinement length scale in the coordinate space, i.e., $r_{\omega}^{} = 1/\omega$. If $\omega \rightarrow 0 $, then $r_{\omega} \rightarrow \infty $ and the interaction model is actually the constant model, which can be expressed as $\delta^4(k)$ in the momentum space~\cite{Munczek:1983PRD}. In such a case, the CEP locates at $({\mu_{E}^{q}}, {T_{E}^{}}) = (0, T_{c}) $, i.e., on the $T$-axis~\cite{Blaschke:1998PLB}. On the contrary, if $\omega \rightarrow \infty$, then $r_{\omega}^{} \rightarrow 0$ and the interaction model becomes the contact interaction model, i.e., the NJL-type model. Many NJL-type calculations have shown that ${\mu_{E}^{q}}/{T_{E}^{}}$ takes large values (see, for example, Refs.~\cite{Jiang:2013PRD,Fukushima:2008PRD,Fu:2008PRD}). Our interaction models are interpolations between these two limits. 
Our results indicate that different methods give distinct locations of the CEP because they have different intrinsic confinement length scales. Recalling the discussion in Ref.~\cite{Xin:2014PRD}, one can understand  the underlying mechanism of such a variation feature as follows. As mentioned above and in Ref.~\cite{Qin:2011PRL}, with decreasing $\omega$, the confinement length between quarks becomes larger, and thus the binding energy of baryons (more generally, hadrons) gets larger. It takes then more energy to release quarks from the color singlet system (DCSB phase or DCSB/DCS coexistence phase), so that $T_{E}^{}$  and $T_{c}^{}$ (for not only the deconfinement but also the chiral phase transitions) get higher as $\omega$ decreases. On the other hand, increasing the confinement length scale plays the same role as increasing the density of the system, it compensates then the effect of increasing the chemical potential. Therefore, $\mu_{E}^{q}$ (also for both the deconfinement and the chiral phase transitions) decreases due to the compensation.

\section{Summary and Remarks}\label{sec4}

Using the Dyson-Schwinger equations approach, we have calculated the quark number fluctuations  up to the forth order in the matter composed of two light flavor quarks with isospin symmetry and at not only finite temperature but also finite chemical potential. The obtained behaviors of the fluctuations at finite temperature and zero chemical potential are consistent with the lattice QCD results. With the MT model and the QC model, we obtain the pseudo-critical temperature $T_c=146$~MeV and $150$~Mev, respectively, which are also consistent with the recent lattice QCD result. Then, we extend our calculations to cases at finite temperature and finite chemical potential. We found that the second order fluctuation exhibits a hump in low chemical potential and high temperature region, and involves singularities in high chemical potential and low temperature region. Meanwhile, the higher order fluctuations oscillate and diverge in the corresponding regions, respectively. These behaviors are the same as those of the chiral susceptibility. So the fluctuations can be taken as a criterion of the phase transition and the location of the CEP. We locate then the CEP at the point with ${\mu_{E}^{q}}/{T_{E}^{}} \sim 1$, which is in nice agreement with the experimental estimate and the lattice QCD results. Moreover, we also find that the experimental quantities which are related to the fluctuation ratios, are very sensitive to the critical behaviors. For searching the CEP, the fluctuations can bridge theories and experiments.

In our calculations, we take two models (the MT model and the QC model) for the dressed gluon propagator. Our obtained results indicate that the fluctuations and the location of the CEP are almost independent of the deep infrared detail of the interaction. We also analyze the dependence of the results on the model parameter. It is found that the location of the CEP depends strongly on the confinement length scale: The CEP rotates to the chemical potential axis with decreasing confinement length scale.

In this work we adopted the rainbow approximation as the first step. Even though the rainbow approximation could partially describe the hadron spectrum, its drawback is also outstanding, e.g., it fails to explain the mass splitting of the vector and axial-vector mesons and also violates the Ward--Green--Takahashi identities. It would then be beneficial to calculate the fluctuations with the Ball-Chiu vertex~\cite{Ball:1980PRD} or the more realistic Anomalous Chromomagnetic Moment (or CLRQ) vertex~\cite{Chang:2011PRL,Qin:2013PLB,Baschir:2012PRC}. Additionally, it would also be interesting to extend our calculations to extract the critical exponents of the fluctuations. The related investigations are under progress.

\bigskip

\begin{acknowledgments}
The work was supported by the National Natural Science Foundation of China under Contract Nos.\ 10935001 and 11175004, and the National Key Basic Research Program of China under Contract No.\ G2013CB834400.
S.-X. Qin thanks also the support of Alexander von Humboldt Foundation.
\end{acknowledgments}


\begin{thebibliography}{100}

\bibitem{Gupta:2011Science}
  S.~Gupta, X.~Luo, B.~Mohanty, H.~G.~Ritter and N.~Xu,
  Science {\bf 332}, 1525 (2011).

\bibitem{Fukushima:2011Review}
      K. Fukushima, and T. Hatsuda,
       Rep. Prog. Phys. {\bf 74}, 014001 (2011).

\bibitem{Wambach:2009Review}
      P. Braun-Munzinger, and J. Wambach,
         Rev. Mod. Phys. {\bf 81}, 1031 (2009).

\bibitem{Shuryak:2004PNPP}
  E.~Shuryak,
  Prog.\ Part.\ Nucl.\ Phys.\  {\bf 53}, 273 (2004).

\bibitem{Melkumov:2012PAN}
   G. L. Melkumov, et al.,
       Nucl. Phys.\ B (Proc. Suppl.)\  {\bf 219}, 102 (2011);
%
   G. L. Melkumov, et al.(NA49 Collaboration),
      Phys.\ At.\ Nucl.\  {\bf 75}, 556 (2012).

\bibitem{Odyniec:2012PAN}
  G. Odyniec,
  Phys.\ At.\ Nucl.\  {\bf 75}, 602 (2012).

\bibitem{Qin:2011PRL}
  S. X. Qin, L. Chang, H. Chen, Y. X. Liu and C. D. Roberts,
  Phys.\ Rev.\ Lett.\  {\bf 106}, 172301 (2011).

\bibitem{Jiang:2013PRD}
  L. J. Jiang, X. Y. Xin, K. L. Wang, S. X. Qin, and Y. X. Liu,
  Phys.\ Rev.\ D {\bf 88}, 016008  (2013).

\bibitem{Liu:20101PRD}
   A. Y. Li, A. Alexandru, K.-F. Liu, and X. Meng,
      Phys. Rev. D {\bf 82}, 054502 (2010);
   A. Y. Li, A. Alexandru, and K.-F. Liu,
      Phys. Rev. D {\bf 84}, 071503(R) (2011).

\bibitem{Stephanov:2009PRL}
  M. A. Stephanov, K. Rajagopal, and E.V. Shuryak,
    Phys. Rev. Lett. {\bf 81}, 4816 (1998);
  M. A. Stephanov,
     Phys. Rev. Lett. {\bf 102}, 032301 (2009).

\bibitem{Stephanov:2011PRL}
   M. A. Stephanov,
     Phys. Rev. Lett. {\bf 107}, 052301 (2011).

\bibitem{Zhou:2010PRC}
   Y. Zhou, S. S. Shi, K. Xiao, K. J. Wu, and F. Liu,
    Phys. Rev. C {\bf 82}, 014905 (2010).

\bibitem{Fraga:2011PRC}
   L. F. Palhares, E. S. Fraga, and T. Kodama,
     J. Phys. G {\bf 37}, 094031 (2010);
     J. Phys. G {\bf 38}, 085101 (2011);
   E. S. Fraga, L. F. Palhares, and P. Sorensen,
    Phys. Rev. C {\bf 84}, 011903(R) (2011).

\bibitem{Kapusta:2012PRC}
   J. I. Kapusta, and J. M. Torres-Rincon,
    Phys. Rev. C {\bf 86}, 054911 (2012).

\bibitem{Stephanov:1999PRD}
  M. A. Stephanov, K. Rajagopal, and E.V. Shuryak,
     Phys. Rev. D {\bf 60}, 114028 (1999).
     J. Phys. G. {\bf 38}, 124147 (2011).

\bibitem{Koch:20058PRLetc}
   V. Koch, A. Majumder, and J. Randrup,
     Phys. Rev. Lett. {\bf 95}, 182301 (2005);
   V.~Koch,
     arXiv:0810.2520 [nucl-th].

\bibitem{Aggarwal:2010PRL}
   M. M. Aggarwal, et al. (STAR Collaboration),
     Phys. Rev. Lett. {\bf 105}, 022302 (2010).

\bibitem{Tarnowsky:2011JPG}
  T.~J. Tarnowsky (for the STAR Collaboration) ,
  J. Phys.\ G\  {\bf 38}, 124054 (2011).

\bibitem{Luo:2012PAN}
   X.~F. Luo, B. Mohanty, H. G. Ritter, and N. Xu,
      Phys.\ At. Nucl.\  {\bf 75}, 676 (2012).
%
   X.~Luo (for the STAR Collaboration),
      PoS CPOD {\bf 2013}, 019 (2013)
      (arXiv:1306.3106 [nucl-ex]).

\bibitem{Mohanty:2013}
   L. Kumar,
     Nucl. Phys. A {\bf 904}, 256c (2013);
%
   R. Singh, L. Kumar, P.K. Netrakanti, and B. Mohanty,
      arXiv:1304.2969;
%
   B. Mohanty,
      arXiv: 1308.3328.

\bibitem{Ejiri:2006PLB}
  S.~Ejiri, F.~Karsch and K.~Redlich,
  Phys.\ Lett.\ B {\bf 633}, 275 (2006).

\bibitem{Cheng:2009PRD}
  M.~Cheng, P.~Hendge, C.~Jung, F.~Karsch, O.~Kaczmarek, E.~Laermann, R.~D.~Mawhinney and C.~Miao {\it et al.},
  Phys.\ Rev.\ D {\bf 79}, 074505 (2009).

\bibitem{Bazavov:2012PRD1}
  A.~Bazavov, T.~Bhattacharya, M.~Cheng, C.~DeTar, H.~T.~Ding, S.~Gottlieb, R.~Gupta and P.~Hegde {\it et al.},
  Phys.\ Rev.\ D {\bf 85}, 054503 (2012).

\bibitem{Bazavov:2012PRD2}
  A.~Bazavov {\it et al.}  [HotQCD Collaboration],
  Phys.\ Rev.\ D {\bf 86}, 034509 (2012).

\bibitem{Aoki:2006PLB}
  Y.~Aoki, Z.~Fodor, S.~D.~Katz and K.~K.~Szab\'{o},
  Phys.\ Lett.\ B {\bf 643}, 46 (2006).

\bibitem{Aoki:2009JHEP}
  Y. Aoki, S. Borsanyi, S. Durr, Z. Fodor, S. D. Katz, S. Krieg,
    and K. K. Szab\'{o},
  JHEP {\bf 06} (2009), 088.

\bibitem{Gavai:2005PRD}
  R. V. Gavai and S. Gupta,
  Phys.\ Rev.\ D {\bf 71}, 114014 (2005).

\bibitem{Gavai:2008PRD}
  R.~V.~Gavai and S.~Gupta,
  Phys.\ Rev.\ D {\bf 78}, 114503 (2008).

\bibitem{Borsanyi:2010JHEP}
   S. Borsanyi, Z. Fodor,  C. Hoelbling, S.D. Katz, S. Krieg, C. Ratti, and K. Szab\'{o},
     J. High Energy Phys. {\bf 09} (2010), 073.
%

\bibitem{Borsanyi:2012JHEP}
   S. Borsanyi, Z. Fodor, S.D. Katz, S. Krieg, C. Ratti, and K. Szab\'{o},
     J. High Energy Phys. {\bf 01} (2012), 138.

\bibitem{Endrodi:2011JHEP}
   G. Endr\"{o}di, Z. Fodor,  S.D. Katz, and K. Szab\'{o},
     J. High Energy Phys. {\bf 04} (2011), 001.

\bibitem{Allton:2003PRD}
  C. R. Allton, S. Ejiri, S. J. Hands, O. Kaczmarek,
  F. Karsch, E. Laermann, and C. Schmidt,
     Phys.\ Rev.\ D {\bf 68}, 014507 (2003).

\bibitem{Allton:2005PRD}
  C. R. Allton, M. Doring, S. Ejiri, S. J. Hands, O. Kaczmarek,
  F. Karsch, E. Laermann and K. Redlich,
     Phys.\ Rev.\ D {\bf 71}, 054508 (2005).

\bibitem{Bernard:20058PRD}
   C.~Bernard {\it et al.}  [MILC Collaboration],
       Phys.\ Rev.\ D {\bf 71}, 034504 (2005);
       Phys. Rev. D {\bf 77}, 014503 (2008).

\bibitem{Karsch:2011PRD}
   F. Karsch, and K. Redlich,
    Phys. Rev. D {\bf 84}, 051504(R) (2011).

\bibitem{Fukushima:2008PRD}
   K. Fukushima,
     Phys. Rev. D {\bf 77}, 114028 (2008).

\bibitem{Fu:2010PRD1}
  W. J. Fu, Y. X. Liu, and Y. L. Wu,
  Phys.\ Rev.\ D {\bf 81}, 014028 (2010).

\bibitem{Fu:2010PRD2}
  W. J. Fu, and Y. L. Wu,
  Phys.\ Rev.\ D {\bf 82}, 074013 (2010).

\bibitem{Ray:20101PRD}
   A. Bhattacharyya, P. Deb, A. Lahiri, and R. Ray,
      Phys. Rev. D {\bf 82}, 114028 (2010);
      Phys. Rev. D {\bf 83}, 014011 (2011).

\bibitem{Schaefer:20102PRD}
   B.-J. Schaefer, M. Wagner, and J. Wambach,
      Phys. Rev. D {\bf 81}, 074013 (2010);
   B.-J. Schaefer, and M. Wagner,
      Phys. Rev. D {\bf 85}, 034027 (2012).

\bibitem{Skokov:2011PRC}
   V. Skokov, B. Friman, and K. Redlich,
      Phys.\ Rev.\ C {\bf 83}, 054904 (2011).

\bibitem{Mohan:2012PRD}
  S. Chatterjee, and K. A. Mohan,
     Phys.\ Rev.\ D {\bf 85}, 074018 (2012);
  S. Chatterjee, and K. A. Mohan,
  Phys.\ Rev.\ D {\bf 86}, 114021 (2012).

\bibitem{Huovinen:2010NPA}
  P.~Huovinen and P.~Petreczky,
  Nucl.\ Phys.\ A {\bf 837}, 26 (2010).

\bibitem{Cleymans:1999PRC}
  J. Cleymans, and K.~Redlich,
  Phys.\ Rev.\ C {\bf 60}, 054908 (1999).

\bibitem{Andronic:2006NPA}
  A. Andronic, P. Braun-Munzinger, and J. Stachel,
  Nucl.\ Phys.\ A {\bf 772}, 167 (2006).

\bibitem{Stokic:2009PLB}
  B. Stokic, B. Friman, and K. Redlich,
  Phys.\ Lett.\ B {\bf 673}, 192 (2009).

\bibitem{Asakawa:2009PRL}
  M. Asakawa, S. Ejiri, and M. Kitazawa,
  Phys.\ Rev.\ Lett.\  {\bf 103}, 262301 (2009).

\bibitem{Roberts:199400DSEint}
   C. D. Roberts, and A. G. Williams,
      Prog. Part. Nucl. Phys. {\bf 33}, 477 (1994);
   C. D. Roberts and S. M. Schmidt,
      Prog. Part. Nucl. Phys. {\bf 45}, S1 (2000).


\bibitem{Zong:2009PRD}
    M. He, J. F. Li, W. M. Sun, and H. S. Zong,
       Phys. Rev. D {\bf 79}, 036001 (2009).

\bibitem{Zong:2009PLB}
  D. K. He, X. X. Ruan, Y. Jiang, W. M. Sun, and H. S. Zong,
     Phys.\ Lett.\ B {\bf 680}, 432 (2009).

\bibitem{Zong:2011JHEP}
  Y. Jiang, L. J. Luo, and H. S. Zong,
     JHEP {\bf 02} (2011), 066.

\bibitem{Alkofer:2001DSE}
   R. Alkofer, and L. von Smekal,
      Phys. Rept. {\bf 353}, 281 (2001);
%
   C. S. Fischer,
      J. Phys. G {\bf  32}, R253 (2006).

\bibitem{Roberts:2003DSE}
   P. Maris and C. D. Roberts,
   Int. J. Mod. Phys. {\bf E 12}, 297 (2003).

\bibitem{Roberts:20123Review}
   A. Bashir, L. Chang, I. C. Cloet, B. El-Bennich, Y. X. Liu, C. D. Roberts, and P. C. Tandy,
      Commun. Theor. Phys. {\bf 58}, 79 (2012);
   C. D. Roberts,
      arXiv:1203.5341 ;
   I.C. Cloet, and C. D. Roberts,
     Prog. Part. Nucl. Phys. {\bf 77}, 1 (2014).

\bibitem{McLerran:2007NPA}
  L.~McLerran and R.~D.~Pisarski,
  Nucl.\ Phys.\ A {\bf 796}, 83 (2007).

\bibitem{Maris:2003EPJA}
   P. Maris, A. Raya, C. D. Roberts, and S. M. Schmidt,
     Eur.\ Phys.\ J.\ A {\bf 18}, 231 (2003).

\bibitem{He:2007JPG}
  M. He, H. T. Feng, W. M. Sun, and H. S. Zong,
     J.\ Phys.\ G {\bf 34}, 2655 (2007).

\bibitem{Qin:2013PRD}
   S. X. Qin, and D. H. Rischke,
      Phys.\ Rev.\ D {\bf 88}, 056007 (2013).

\bibitem{Fischer:2009PRL}
    C. S. Fischer,
       Phys. Rev. Lett. {\bf 103}, 052003 (2009);
%
    C. S. Fischer, and J. A. Mueller,
       Phys. Rev. D {\bf 80}, 074029 (2009).

\bibitem{Fischer:20113PLB}
    C. S. Fischer, J. Luecker, J. A. Mueller,
       Phys. Lett. B {\bf 702}, 438 (2011);
%
    C. S. Fischer, and J. Luecker,
       Phys. Lett. B {\bf 718}, 1036 (2013).

\bibitem{Qin:2011PRD}
   J. A. Mueller, C. S. Fischer, and D. Nickel,
      Eur. Phys. J. C {\bf 70}, 1037 (2010);
   S. X. Qin, L. Chang, Y. X. Liu, and C. D. Roberts,
      Phys.\ Rev.\ D {\bf 84}, 014017 (2011).
%
   F. Gao, S. X. Qin, Y. X. Liu, C. D. Roberts, and S. M. Schmidt,
      Phys. Rev. D {\bf 89}, 076009 (2014).

\bibitem{Chang:2009PRL}
   L. Chang, and C. D. Roberts,
     Phys. Rev. Lett.  {\bf 103}, 081601 (2009);
     Phys. Rev. C {\bf 85}, 052201(R) (2012).

\bibitem{Qin:2011PRC}
   S. X. Qin, L. Chang, Y. X. Liu, C. D. Roberts, and D. J. Wilson,
      Phys.\ Rev.\ C {\bf 84}, 042202(R) (2011).

\bibitem{Qin:2012PRC}
  S. X. Qin, L. Chang, Y. X. Liu, C. D. Roberts, and D. J. Wilson,
      Phys.\ Rev.\ C {\bf 85}, 035202 (2012).

\bibitem{Eichmann:2010PRL}
   G. Eichmann, I. C. Cloet, R. Alkofer, A. Krassnigg, and C. D. Roberts,
      Phys. Rev. C {\bf 79}, 012202(R) (2009);
%
   G. Eichmann, R. Alkofer, A. Krassnigg, and D. Nicmorus,
      Phys. Rev. Lett. {\bf 104}, 201601 (2010).

\bibitem{Eichmann:2011PRD}
   I. C. Cloet, G. Eichmann, B. El-Bennich, T. Klaehn, and C. D. Roberts,
      Few-Body Systems {\bf 46}, 1 (2009);
%
   G. Eichmann,
      Phys. Rev. D {\bf 84}, 014014 (2011);
%
   G. Eichmann, and C. S. Fischer,
     Eur. Phys. J. A {\bf 48}, 1 (2012).

\bibitem{Chang:2013PRL}
   L. Chang, C. D. Roberts, and S. M. Schmidt,
      Phys. Rev. C {\bf 87}, 015203 (2013);
%
   L. Chang, I. C. Cloet, J. J. Cobos-Martinez, C. D. Roberts, S. M. Schmidt, and P. C. Tandy,
      Phys. Rev. Lett. {\bf 110}, 132001 (2013);
%
   I. C. Cloet, L. Chang, C. D. Roberts, S. M. Schmidt, and P. C. Tandy,
      Phys. Rev. Lett. {\bf 111}, 092001 (2013);
%
   L. Chang, I. C. Cloet, C. D. Roberts, S. M. Schmidt, and P. C. Tandy,
      Phys. Rev. Lett. {\bf 111}, 141802 (2013);
%
   C. D. Roberts, R. J. Holt, and S. M. Schmidt,
      Phys. Lett. B {\bf 727}, 249 (2013).

\bibitem{Wang:2013PRD}
   K. L. Wang, Y. X. Liu, L. Chang, C. D. Roberts, and S. M. Schmidt,
     Phys. Rev. D. {\bf 87}, 074038 (2013).

\bibitem{Maris:1999PRC}
  P.~Maris and P.~C.~Tandy,
  Phys.\ Rev.\ C {\bf 60}, 055214 (1999).

\bibitem{Oliveira:2011JPG}
   O.~Oliveira and P.~Bicudo,
      J.\ Phys.\ G {\bf 38}, 045003 (2011);
   O.~Oliveira, and P. J. Silva,
      Phys. Rev. D {\bf 86}, 114513 (2012).

\bibitem{Aguilar:2012PRD}
  A. C. Aguilar, D. Binosi, and J. Papavassiliou,
  Phys.\ Rev.\ D {\bf 86}, 014032 (2012).

\bibitem{Fischer:2012PRL}
   S. Strauss, C. S. Fischer, and C. Kellermann,
       Phys. Rev. Lett. {\bf 109}, 252001 (2012).

\bibitem{Dudal:2012PRD}
   A. Cucchieri, D. Dudal, T. Mendes and N. Vandersickel,
     Phys. Rev. D {\bf 85}, 094513 (2012);
   A. Ayala, A. Bashir, D. Binosi, M. Cristoforetti and
      J. Rodriguez-Quintero,
       Phys. Rev. D {\bf 86}, 074512 (2012);
    D. Dudal, O. Oliveira and J. Rodriguez-Quintero,
      Phys. Rev. D {\bf 86}, 105005 (2012);
   B. Blossier, Ph. Boucaud, M. Brinet, F. De Soto, V. Morenas,
   O. P\'{e}ne, K. Petrov, and J. Rodriguez-Quintero,
      Phys. Rev. D {\bf 87}, 074033 (2013).

\bibitem{Zwanziger:2013PRD}
   D. Zwanziger,
      Phys. Rev. D {\bf 87}, 085039 (2013).

\bibitem{Stephanov:2004APPB}
  M. Stephanov,
  Acta Phys.\ Polon.\ B {\bf 35}, 2939 (2004).

\bibitem{Luecker:2013PoS}
  J.~Luecker, C.~S.~Fischer, L.~Fister and J.~M.~Pawlowski,
  PoS CPOD {\bf 2013}, 057 (2013).

\bibitem{Kondo:1997IJMPA}
  K. I. Kondo,
  Int.\ J.\ Mod.\ Phys.\ A {\bf 12}, 5651 (1997).

\bibitem{Karsch:2009PRD}
   F. Karsch, and M. Kitazawa,
     Phys. Rev. D {\bf 80}, 056001 (2009).

\bibitem{Maris:1997PRC}
  P.~Maris and C.~D.~Roberts,
  Phys.\ Rev.\ C {\bf 56}, 3369 (1997).

\bibitem{Hawes:1994PRD}
  F.~T.~Hawes, C.~D.~Roberts and A.~G.~Williams,
  Phys.\ Rev.\ D {\bf 49}, 4683 (1994).

\bibitem{Rajagopal:2000PRD}
    B. Berdnikov, and K. Rajagopal,
       Phys. Rev. D {\bf 61}, 105017 (2000).

\bibitem{Lacey:2007na}
   R. A. Lacey, N. N. Ajitanand, J. M. Alexander, P. Chung, J. Jia, A. Taranenko and P. Danielewicz,
     arXiv:0708.3512 [nucl-ex].

\bibitem{Weise:200910PRD}
   T. Hell, S. Roessner, M. Cristoforetti, and W. Weise,
     Phys. Rev. D {\bf 79}, 014022 (2009);
     Phys. Rev. D {\bf 81}, 074034 (2010).

\bibitem{Contrera:2010}
   G. A. Contrera, M. Orsaria, and N. N. Scoccola,
     Phys. Rev. D {\bf 82}, 054026 (2010).


\bibitem{Fodor:2002JHEP}
     Z. Fodor, and S. D. Katz,
        J. High Energy Phys. {\bf 03} (2002), 014.
%

\bibitem{Schmidt:2008JPG}
     C. Schmidt (for the RBC-Bielefeld Collaboration),
        J. Phys. G {\bf 35}, 104093 (2008).
%

\bibitem{Fodor:2004JHEP}
     Z. Fodor, and S. D. Katz,
        J. High Energy Phys. {\bf 04} (2004), 050.
%

\bibitem{Gupta:2014}
     S. Gupta, N. Karthik, and P. Majumdar,
        arXiv:1405.2206 [hep-lat].
%

\bibitem{Munczek:1983PRD}
  H. J. Munczek, and A. M. Nemirovsky,
  Phys.\ Rev.\ D {\bf 28}, 181 (1983).

\bibitem{Blaschke:1998PLB}
  D. Blaschke, C. D. Roberts, and S. M. Schmidt,
  Phys.\ Lett.\ B {\bf 425}, 232 (1998).

\bibitem{Fu:2008PRD}
  W. J. Fu, Z. Zhang, and Y. X. Liu,
  Phys.\ Rev.\ D {\bf 77}, 014006 (2008).

\bibitem{Xin:2014PRD}
   X. Y. Xin, S. X. Qin, and Y. X. Liu,
  Phys. Rev. D {\bf 89}, 094012 (2014).

\bibitem{Ball:1980PRD}
  J.~S.~Ball and T.~-W.~Chiu,
  Phys.\ Rev.\ D {\bf 22}, 2542 (1980).

\bibitem{Chang:2011PRL}
   L. Chang, Y. X. Liu, and C. D. Roberts,
     Phys.\ Rev.\ Lett.\  {\bf 106}, 072001 (2011).

\bibitem{Qin:2013PLB}
  S. X. Qin, L. Chang, Y. X. Liu, C. D. Roberts and S. M. Schmidt,
     Phys.\ Lett.\ B {\bf 722}, 384 (2013);
%
  S. X. Qin, C. D. Roberts, and S. M. Schmidt,
      Phys. Lett. B {\bf 733}, 202 (2014).

\bibitem{Baschir:2012PRC}
   A. Bashir, R. Bermudez, L. Chang, C. D. Roberts,
     Phys. Rev. C {\bf 85}, 045205 (2012).

\end{thebibliography}


\end{document}